\documentclass[11pt]{article}

\usepackage{amsmath}
\usepackage{amssymb}
\usepackage{epsfig}

\begin{document}


\renewcommand{\thefootnote}{\alph{footnote}}
\vspace*{-3.cm}
\begin{flushright}

\end{flushright}

\vspace*{0.5cm}

\renewcommand{\thefootnote}{\fnsymbol{footnote}}
\setcounter{footnote}{-1}

{\begin{center} {\Large\bf Comment on "Remark on the renormalization
group equation for the Penner model"}

\end{center}}
\renewcommand{\thefootnote}{\alph{footnote}}

\vspace*{.8cm} {\begin{center} {\large{\sc
                Noureddine~Chair
                }}
\end{center}}
\vspace*{0cm} {\it
\begin{center}
 Physics Department,
Al al-Bayt University, Mafraq, Jordan

Email: n.chair@rocketmail.com\\
\hspace{19mm}nchair@aabu.edu.jo
\end{center} }
\vspace*{1.5cm}

{\Large \bf
\begin{center} Abstract\end{center} }
\ We show explicitly that the sum over punctures for the three times
derivative for the Penner free energy $F_{0}^{3}$, given by D.A.
Johnston , Phys.Rev.D 51 (1995) is not correct. As a consequence,
Eq.(21), the differentiated version for the renormalization group
(RG) equation, is wrong. Also, his conclusion  that the
differentiated version of the (RG) equation for the three-times
derivative of the free energy can be obtained from the higher genus
(RG) equation can not be true. Finally, the differentiated version
of the (RG) equation is extended  to any $s$ derivative of the free
energy $F_{0}$.\vspace*{.5cm}

\newpage
\renewcommand{\thefootnote}{\arabic{footnote}}
\setcounter{footnote}{0}


%
\ In \cite{johnston}Phys.Rev.D 51 (1995) 2014 D.A. Johnston derived
the renormalization group of the Penner model\cite{penner}, and
showed to be of the form of the matrix model studied in
\cite{higuchi}.  By using the renormalization group equation, he
showed that the zero of the $\beta$ function, coincides with the
critical coupling constant of the Penner model, this coupling is
known to be $t=-1$ \cite{distler}.  Instead of using the topological
expansion of the free energy given by Distler and Vafa
\cite{distler}, one uses the free energy  given by Penner
\cite{penner}; then the critical coupling constant is $t=1$,
\cite{chair}.  This is simply because the two topological expansions
of the free energy are related by the transformation
$t\rightarrow-\frac{1}{t}$, in the coupling constant. The scaling
exponents that occur in the double scaling limit of the free energy,
for the $c=1$ matrix model \cite{brezin}, are also computed from the
renormalization group equation. This shows that the Penner model is
a $c=1$ matrix model. In these comments, we show that the Eqs .(16),
(20) and (21) given by Johnston in his paper are not correct.  Also,
his conclusion  that the differentiated version of the
renormalization group equation (RG) for the three-times derivative
of the free energy $F_{0}^{3}$ can be obtained from the higher genus
(RG) equation, [see Eq. (13)] can not be true.

\ These mistakes are easily found by writing the Penner free energy
\cite{distler} and using the constraint on the minimum number of
punctures that can be put on a given Riemann surface of genus $g$.
Recall that the Penner free energy is
\begin{equation}
\label{1}
F(t,N) = \sum_{g,n} N^{2 - 2 g} t^{2-n-2g } \chi_{g,n},
\end{equation}
where the coefficients $\chi_{g,s}$ are the orbifold Euler
characteristics of the moduli space of Riemann surfaces of genus $g$
with $n$ punctures; explicitly, this topological invariant is given
by
\begin{equation}
\label{2}
\chi_{g,n} = { (-1)^{n}( 2 g - 3 + n ) ! ( 2 g - 1 ) \over
( 2 g ) ! n ! } B_{2g}
\end{equation}
where $B_{2g}$ are the Bernoulli numbers.  If we write $F(t,N) =
\sum_{g} N^{2 - 2 g}F_{g}$, where $F_{g}=\sum_{n}t^{2-n-2g }
\chi_{g,n}$, then it is true that the sum over punctures starts from
$n=1$, but only for genus $g\geq1$ \cite{penner}.  For $g=0$, the
sum should start from $n=3$; this is clear from equation (\ref{2}).
Therefore the $g=0$ free energy $F_{0}$ is given by
\begin{equation}
\label{3} F_{0}=\sum_{n\geq3}(-1)^{n+1}\frac{( n-3 )!}{ n!}~t^{2-n};
\end{equation}
hence, the three-times derivative of $F_{0}$, with respect to $t$ is
\begin{equation}
\label{4}
F_{0}^{3}= \sum_{n\geq3} (-1)^{n}t^{2-n}=
-\frac{N^{2}}{t^{3}(1+t)}.
\end{equation}
This differs from equation equation (16), and hence from equation
(20), where the sum starts  from $n=1$, which is not allowed.

\ Using the expression for the free energy Eq.(\ref{3}), the
analogue of the (RG) equation for $g=0$, given by Johnston, Eq.
(13) is
\begin{equation}
\label{5}
F_{0}-(1+t)\frac{\partial F_{0}}{\partial
t}=\frac{N^{2}}{2t^{2}}.
\end{equation}
It is interesting to note that by differentiating  Eq. (\ref{5})
twice, one obtains  $F_{0}^{3}$, given by Eq. (\ref{4}), and we do
not have to sum over the punctures. And by differentiating Eq.
(\ref{5}) three times, we obtain the differentiated version of the
(RG) equation for $F_{0}^{3}$,
\begin{equation}
\label{6}
F_{0}^{3}+(1+t)\frac{\partial F_{0}^{3}}{\partial
t}=\frac{3N^{2}}{t^{4}}.
\end{equation}
This equation, of course, can not be obtained from the (RG) equation
(13), given by Johnston, since in obtaining this equation, the sum
over the punctures in the free energy starts from one puncture and
higher.  We would like to add that the differentiated version of the
(RG) equation for $F_{0}^{3}$ can be extended easily to any $s$
derivative of the free energy $F_{0}$, and we obtain
\begin{equation}
\label{7}
(s-2)F_{0}^{s}+(1+t)\frac{\partial F_{0}^{s}}{\partial
t}=\frac{N^{2}(-1)^{s+1}s!}{2t^{s+1}}.
\end{equation}
As one can see, that this differentiated version of the (RG)
equation for $F_{0}^{s}$ is generally true for all $s$;  in
particular, for $s=0$, $s=3$ gives equations (\ref{5}), (\ref{6})
respectively.
\newpage

\bibliographystyle{phaip}

\end{document}